\documentclass[prd,preprint,nofootinbib]{revtex4}
\usepackage{amssymb}
\usepackage{epsfig}

\begin{document}
\title{ 
 Light W-ino dark matter in brane world cosmology
}

\author{Takeshi Nihei}
 \email{nihei@phys.cst.nihon-u.ac.jp}
 \affiliation{
Department of Physics, College of Science and Technology, 
Nihon University, \\ 
1-8-14, Kanda-Surugadai, Chiyoda-ku, Tokyo 101-8308, Japan
 }

\author{Nobuchika Okada}
 \email{okadan@post.kek.jp}
 \affiliation{
Theory Division, KEK, Oho 1-1, Tsukuba, Ibaraki 305-0801, Japan \\ 
Department of Particle and Nuclear Physics, The Graduate University \\ 
 for Advanced Studies (Sokendai), 
 Oho 1-1, Tsukuba, Ibaraki 305-0801, Japan 
 }

\author{Osamu Seto}
 \email{O.Seto@sussex.ac.uk}
 \affiliation{
  Department of Physics and Astronomy, University of Sussex, 
  Brighton BN1 9QJ, United Kingdom
 }


\begin{abstract}
The thermal relic density of the W-ino-like neutralino dark matter 
 in brane world cosmology is studied. 
The expansion law at a high energy regime 
 in brane world cosmology is modified 
 from the one in standard cosmology, 
 and the resultant relic density can be enhanced 
 if the five-dimensional Planck mass $M_5$ is low enough. 
We calculate the W-ino-like neutralino relic density 
 in the anomaly mediated supersymmetry breaking scenario 
 and show that the allowed region is dramatically modified 
 from the one in standard cosmology 
 and the W-ino-like neutralino with mass of order 100 GeV 
 can be a good candidate for dark matter. 
Since the allowed region disappears eventually as $M_5$ decreases, 
 we can find a lower bound on $M_5 \gtrsim 100$ TeV 
 according to the neutralino dark matter hypothesis, 
 namely the lower bound in order for the allowed region 
 of neutralino dark matter to exist. 
\end{abstract}

\preprint{KEK-TH-1032}

\vspace*{1cm}

\maketitle


\section{INTRODUCTION}

Recent cosmological observations,
 especially after the Wilkinson Microwave Anisotropy Probe (WMAP) 
 satellite \cite{WMAP}, 
 have established the $\Lambda$CDM cosmological model
 with a great accuracy.
The abundance of cold dark matter is also precisely measured as
 (in 2 $\sigma$ range) 
\begin{equation}
\Omega_{CDM} h^2 = 0.1126^{+0.0161}_{-0.0181} .
\label{OmegaCDM}
\end{equation}
However, to clarify the identity of a particle as cold dark matter 
 is still an open problem in cosmology and particle physics. 

The lightest supersymmetric particle (LSP) in supersymmetric models
 is suitable for cold dark matter, 
 because it is stable owing to the conservation of R parity.
In the minimal supersymmetric standard model (MSSM), 
 the lightest neutralino is typically the LSP 
 and the promising candidate for cold dark matter. 
Neutralinos consist of a mixture of neutral gauginos and Higgsinos,
\begin{equation}
\chi^0_i = N_{i1}\tilde{B}+N_{i2}\tilde{W}^3+N_{i3}\tilde{H}_1^0
 +N_{i4}\tilde{H}_2^0,
\end{equation}
 where $\tilde{B}$ is the $U(1)_Y$ gaugino (b-ino), 
 $\tilde{W}^3$ is the $SU(2)_L$ gaugino (W-ino), and 
 $\tilde{H}_1^0$ and $\tilde{H}_2^0$ are two neutral Higgsinos.
The components of the neutralinos $\chi^0_i$,
 in other words the coefficients $N_{ij}$, 
 are model dependent. 

For example, in the constrained MSSM (CMSSM), the lightest neutralino is 
 b-ino-like $\chi^0_1 \sim \tilde{B}$ in most of the parameter region. 
The exception is the so-called ``focus point'' region 
 where the lightest neutralino is Higgsino-like.
In the light of the WMAP data, the parameter space in the CMSSM 
 which allows the neutralino relic density 
 suitable for cold dark matter has been reanalyzed 
 and the allowed region is dramatically reduced 
 due to the great accuracy of the WMAP data \cite{CDM}. 
The main reason is that
 the predicted relic density of the b-ino-like neutralino is generally too much 
 to meet the observational result
 because its annihilation cross section is too small.

However, in another model,
 the lightest neutralino can be W-ino-like or Higgsino-like.
For example, the W-ino-like lightest neutralino $\chi^0_1 \sim \tilde{W}^3$ 
 is naturally realized
 in the anomaly mediated supersymmetry breaking (AMSB) scenario 
 \cite{AnomalyMediation1, AnomalyMediation2}.
The annihilation cross section of the W-ino-like neutralino is
 larger than that of the b-ino-like neutralino, 
 since its pair annihilation process into two W bosons 
 and coannihilation process with charginos 
 via W boson exchange are governed by the $SU(2)_L$ gauge coupling. 
The predicted relic density of the W-ino-like neutralino 
 is roughly estimated as \cite{AnomalyMediation2, Mizuta} 
\begin{equation}
\Omega_{\tilde{W}}h^2 \simeq 5 \times
 10^{-4}\left(\frac{m_{\tilde{W}}}{100 \textrm{GeV}}\right)^2. 
\label{OmegaWino}
\end{equation}
Hence, the W-ino mass $m_{\tilde{W}}$ around $1$ TeV is necessary 
 to be consistent with the WMAP data, 
 and the W-ino-like neutralino seems not to be an appealing candidate 
 in the viewpoint of the weak scale supersymmetry. 
There are some possibilities to make the W-ino-like neutralino 
 with mass of $\mathcal{O}(100)$ GeV a reasonable dark matter candidate, 
 such as nonthermal production of the W-ino-like neutralino 
 \cite{MoroiRandall, FujiiHamaguchi, Pallis} 
 or the introduction of additional energy component 
 at the neutralino decoupling time \cite{AdditionalScalar}. 

Here, we remind the reader that the thermal relic density of dark matter 
 depends on the underlying cosmological model 
 as well as its annihilation cross section. 
If a non-standard cosmology has been realized 
 at the neutralino decoupling time, 
 the resultant relic density of the dark matter 
 can be altered from the one in standard cosmology. 
A brane world cosmological model that is
 being intensively investigated \cite{braneworld} 
 is an interesting example 
 which provide an unconventional cosmological expansion 
 in the early universe. 
In the cosmology based on the so-called ``RS II'' model, 
 first proposed by Randall and Sundrum \cite{RS}, 
 the Friedmann equation for a spatially flat spacetime 
 is found to be 
\begin{equation}
H^2 = \frac{8\pi G}{3}\rho\left(1+\frac{\rho}{\rho_0}\right),
\label{BraneFriedmannEq}
\end{equation}
 where
\begin{eqnarray}
\rho_0 = 96 \pi G M_5^6,
\label{def:rho_0}
\end{eqnarray}
 $H$ is the Hubble parameter, $\rho$ is the energy density of matter, 
 $G$ is the Newton's gravitational constant 
 with $M_5$ being the five-dimensional Planck mass, 
 and the four-dimensional cosmological constant has been 
 tuned to be almost zero. 
The parameter $\rho_0$ is constrained 
 by big bang nucleosynthesis (BBN), 
 which is roughly given by $\rho_0^{1/4} \gtrsim 1$ MeV 
 (or equivalently $M_5 \gtrsim 8.8$ TeV) \cite{braneworld}.
This is a model-independent cosmological constraint. 
On the other hand, 
 as discussed in the original paper by Randall and Sundrum \cite{RS}, 
 the precision measurements of the gravitational law 
 in the submillimeter range lead to more stringent constraint 
 $\rho_0^{1/4} \gtrsim 1.3$ TeV 
 (or equivalently $M_5 \gtrsim 1.1 \times 10^8$ GeV)  
 through the vanishing cosmological constant condition. 
However note that this constraint, in general, 
 is quite model dependent. 
In fact, if we consider an extension of the model 
 so as to introduce a bulk scalar field, 
 the constraint can be moderated because of the change of 
 the vanishing cosmological constant condition \cite{maedawands}. 
Hence, we care about only the BBN constraint 
 on $\rho_0$ in this paper. 

The $\rho^2$ term in Eq.~(\ref{BraneFriedmannEq}) 
 is a new ingredient in brane world cosmology 
 and it dominates over the linear term at a high temperature regime. 
This modification of the expansion law 
 can lead some drastic changes for some results 
 previously obtained in standard cosmology 
 \cite{OS, NOS, OSGravitino, Axino, OSleptogenesis}. 
In particular, it has been pointed out that 
 the thermal relic abundance of dark matter 
 can be considerably enhanced 
 compared to that in standard cosmology \cite{OS}. 
When this scenario is applied to 
 the analysis of the neutralino dark matter in the CMSSM, 
 the allowed parameter region is dramatically modified 
 and eventually disappears 
 as the five-dimensional Planck mass ($M_5$) becomes small,  
 and thus the lower bound on $M_5$ can be obtained \cite{NOS}. 
On the other hand, note that this enhancement implies that 
 a light W-ino dark matter would be favored 
 in brane world cosmology 
 because its annihilation processes are so efficient 
 that its relic density is found to be too small 
 in standard cosmology. 

In this paper, 
 we investigate the possibility that W-ino-like neutralino is 
 a suitable candidate for cold dark matter
 with the help of the brane world cosmological effect. 
This paper is organized as follows. 
In the next section, we give a brief review of Ref.~\cite{OS}. 
In sec.~III, we present our numerical results 
 for the W-ino-like neutralino relic density in brane world cosmology, 
 which shows that the cosmologically interesting parameter region 
 for the W-ino-like neutralino actually appears. 
Sec.~IV is devoted to conclusions.

\section{Enhancement of relic density in brane world cosmology}

In this section, we give a brief review
 on the relic density of dark matter
 in brane world cosmology with 
 a low five-dimensional Planck mass $M_5$ \cite{OS}
 and roughly estimate the suitable scale of $M_5$ for W-ino dark matter. 

In the context of brane world cosmology, 
 we estimate the thermal relic density of 
 a dark matter particle by solving the Boltzmann equation
\begin{equation}
\frac{d n}{d t}+3Hn = -\langle\sigma v\rangle(n^2-n_{\rm{eq}}^2),
\label{n;Boltzmann}
\end{equation}
 with the modified Friedmann equation Eq.~(\ref{BraneFriedmannEq}),
 where $n$ is the actual number density of the dark matter particles,
 $n_{\rm{eq}}$ is the equilibrium number density,
 $\langle\sigma v\rangle$ is the thermal averaged product
 of the annihilation cross section $\sigma$ and the relative velocity $v$.
It is useful to rewrite Eq.~(\ref{n;Boltzmann}) into the form,  
\begin{eqnarray}
\frac{d Y}{d x}
&=& -\frac{s}{xH}\langle\sigma v\rangle(Y^2-Y_{\rm{eq}}^2) \nonumber\\
&=& -\lambda\frac{x^{-2}}{\sqrt{1+ \left(\frac{x_t}{x} \right)^4}}
\langle\sigma v\rangle(Y^2-Y_{\rm{eq}}^2) ,
\label{Y;Boltzmann}
\end{eqnarray}
in terms of the number density to entropy ratio $Y = n/s$ and $x = m/T$, 
 where $m$ is the mass of a dark matter particle, 
 $\lambda = 0.26 (g_{*S}/g_*^{1/2}) M_{P} m$, 
 $M_P \simeq 1.2 \times 10^{19}$GeV is the Planck mass 
 and $x_t$ is defined as 
\begin{equation}
 x_t^4 \equiv \left. \frac{\rho}{\rho_0}\right|_{T=m}. 
\label{def:xt}
\end{equation}
At the era $x \ll x_t$ the $\rho^2$ term dominates 
 in Eq.~(\ref{BraneFriedmannEq}), 
 while the $\rho^2$ term becomes negligible after $x \gg x_t$ 
 and the expansion law in standard cosmology is realized. 
Hereafter we call the temperature defined 
 as $T_t = m x_t^{-1}$ (or $x_t$ itself) ``transition temperature''
 at which the expansion law of the early universe changes 
 from the nonstandard one to the standard one. 
Since we are interested in the effect of the $\rho^2$ term 
 for the dark matter relic density, 
 we consider the case that the decoupling temperature 
 of dark matter ($T_d$) is higher than the transition temperature, 
 namely $x_t \geq x_d=m/T_d$. 

Although it is easy to numerically solve 
 the Boltzmann equation Eq.~(\ref{Y;Boltzmann}) 
 given $\langle \sigma v \rangle$ and $x_t$ 
 as will be shown in the next section,
 here we derive analytic formulas 
 for the relic number density of dark matter 
 by adopting appropriate approximations. 
When we parameterize $\langle \sigma v \rangle$ 
 as $\langle \sigma v \rangle = \sigma_n x^{-n}$ 
 with fixed $n=0,1,\cdots$, for simplicity, 
 we can obtain simple formulas 
 for the resultant relic densities 
 such that 
\begin{eqnarray}
 Y(x \rightarrow \infty) 
 &\simeq& 0.54 \frac{x_t}{\lambda \sigma_0} 
 \qquad \textrm{for} \quad n=0, \nonumber \\
 && \frac{x_t^2}{\lambda \sigma_1 \ln x_t} \qquad \textrm{for} \quad n=1, 
\label{Ybrane}
\end{eqnarray}
in the limit $x_d \ll x_t$, 
 where $x_d$ is the decoupling temperature \cite{OS}. 
Note that the results are characterized 
 by the transition temperature 
 rather than the decoupling temperature. 
It is interesting to compare these results 
 to that in standard cosmology. 
Using the well-known approximate formulas 
 in the standard cosmology \cite{Kolb},  
\begin{eqnarray}
Y(x \rightarrow \infty) 
 &\simeq & \frac{x_d}{\lambda \sigma_0} 
 \qquad \textrm{for} \quad n=0, \nonumber \\ 
 && \frac{2 x_d^2}{\lambda \sigma_1} 
 \qquad \textrm{for} \quad n=1,  
\label{Ystandard}
\end{eqnarray} 
we obtain the ratio of the relic energy density of dark matter 
 in brane world cosmology ($\Omega_{(b)}$) 
 to the one in standard cosmology ($\Omega_{(s)}$) 
 such that 
\begin{eqnarray}
\frac{\Omega_{(b)}}{\Omega_{(s)}} 
 &\simeq& 0.54 \left(  \frac{x_t}{x_{d (s)}}  \right) 
 \qquad \textrm{for} \quad n=0, \label{OmegaRatio} \\ 
 && \frac{1}{2 \ln x_t} 
    \left(  \frac{x_t}{x_{d (s)}}  \right)^2 
 \qquad \textrm{for} \quad n=1, 
\end{eqnarray}
where $x_{d (s)}$ denotes the decoupling temperature 
 in standard cosmology. 
Thus, the relic energy density in brane world cosmology 
 can be enhanced from the one in standard cosmology 
 if the transition temperature is low enough. 

Together with Eqs. (\ref{OmegaCDM}), (\ref{OmegaWino}) 
 and (\ref{OmegaRatio}), we find 
\begin{eqnarray}
 {\Omega_{(b)}} h^2 \simeq 0.1 \times 
 \left( \frac{m}{100 \textrm{GeV}} \right)^2  
 \left( \frac{27}{x_{d (s)}} \right) 
 \left( \frac{x_t}{10^4}  \right)  .
\end{eqnarray}
Therefore, the enhancement by the modified expansion law 
 can make the W-ino-like neutralino with mass ${\cal O}$(100 GeV) 
 a suitable dark matter candidate 
 for $ ( m/100 \textrm{GeV})^2  (x_t/10^4) \simeq 1 $; 
 in other words, 
\begin{equation}
 T_t \simeq  10 \textrm{MeV} 
 \left(\frac{m}{100 \textrm{GeV}}\right)^3,  
\end{equation}
 which corresponds to 
\begin{equation}
 M_5 \simeq 7.3 \times 10^4 \; 
 \left( \frac{g_*}{100} \right)^{1/6} \; 
 \left( \frac{m}{100 \textrm{GeV}} \right)^2  \; \textrm{GeV} 
\end{equation}
 from Eq. (\ref{def:xt}). 
Thus, in brane world cosmology with $M_5$ as low as 100 TeV, 
 the W-ino-like neutralino can be a suitable candidate 
 for the cold dark matter 
 even if its mass scale is of order 100 GeV. 
On the other hand, too small $M_5$ causes the overproduction of dark matter 
 in the universe, and the lower bound on $M_5$ can be obtained 
 for a fixed $m ={\cal O}$(100) GeV.

\section{Numerical results for the wino-like neutralino}

In this section, we present the results of our numerical analysis. 
We calculate the relic density of the W-ino-like neutralino
 $\Omega_{\chi}h^2$ based on the mass spectrum predicted 
 in AMSB scenario to realize the W-ino-like lightest neutralino, 
 with the new Friedmann equation Eq.~(\ref{BraneFriedmannEq}). 
For this purpose, 
 we have modified the code {\sc DarkSUSY} \cite{ref:darksusy} 
 so that the new Friedmann equation is implemented. 
In evaluating the relic density, the relevant coannihilations
 are taken into account. 
In addition, the spin-independent cross section for the direct detection is 
 also estimated.

The mass spectra in the AMSB scenario are determined by 
 the following input parameters \cite{ref:minimal-AMSB}, 
\begin{eqnarray}
 m_0, \ \ m_{3/2}, \ \ \tan\beta, \ \ {\rm sgn}(\mu), 
 \label{eqn:mssm-parameters}
\end{eqnarray}
where $m_{3/2}$ is the gravitino mass, 
 $m_0$ is the universal scalar mass (which is necessary to avoid
 tachyonic slepton mass problem), 
 $\tan\beta$ is the ratio of the vacuum expectation values 
 of the two neutral Higgs fields,
 and ${\rm sgn}(\mu)$ is 
 the sign of the Higgsino mass parameter $\mu$. 
With these input parameters, 
 renormalization group equations for the soft supersymmetry breaking
 parameters are solved using the code {\sc Suspect} \cite{ref:suspect} 
 to obtain the mass spectra at the weak scale. 

In the AMSB scenario, the gaugino mass $M_i$ is proportional to 
 the associated beta function, and each gaugino masses 
 at the weak scale are predicted 
as follows \cite{ref:minimal-AMSB}\footnote{
In the AMSB scenario, gravitino is very heavy 
 and decays before the BBN era, 
 so that the well-known gravitino problem can be avoided. 
In addition, as pointed out in Ref.~\cite{OSGravitino}, 
 there is no gravitino problem in the brane world scenario 
 with the transition temperature low enough $T_t \lesssim 10^6$ GeV, 
 even if gravitino mass is around 100 GeV. 
}: 
\begin{eqnarray}
M_1 & = & 8.9 \times 10^{-3} m_{3/2}, \\
M_2 & = & 2.7 \times 10^{-3} m_{3/2}, \\
M_3 & = & -2.6 \times 10^{-2} m_{3/2} .
\end{eqnarray}
The relation $M_2 \simeq 0.3 M_1$ implies that 
 the lightest neutralino is W-ino-like. 
The lighter chargino is also W-ino-like, 
 and highly degenerated with the lightest neutralino.

\subsection{Relic density of the W-ino-like neutralino}

Figures 1-3 show the allowed region (shaded area) 
for $\tan\beta=10$ and $\mu>0$ in the ($m_{3/2},\, m_0$) 
 plane consistent with the WMAP 2$\sigma$ allowed range 
 $0.094 < \Omega_{\chi} h^2 < 0.129$.
The five-dimensional Planck mass is taken as 
$M_5$ $=$ $\infty$, 500 TeV, and 100 TeV in Figs. 1-3, respectively.

Figure 1 corresponds to the result for standard cosmology. 
In this figure, 
the dotted, dot--long-dashed, short-dash--long-dashed lines 
 correspond to $\Omega_{\chi} h^2$ $=$ 0.005, 0.001 and 0.0005, respectively.
The region consistent with the WMAP data appears 
 in the range $m_{3/2} \simeq 650-850$ TeV, 
 which corresponds to the wino-like neutralino mass around 2 TeV. 
The relic density primarily depends on $m_{3/2}$, and is insensitive to
 $m_0$ as expected from Eq. (\ref{OmegaWino}). 
The region along the bold line and the two coordinate axes 
 is excluded by various experimental constraints 
 (the lightest Higgs mass bound, $b\to s\gamma$ constraint, 
 the lighter chargino mass bound, etc.)
 or the condition for the successful electroweak symmetry breaking. 
In particular, 
 the region $m_0 \lesssim 5 \times 10^{-3} m_{3/2}$ is excluded 
 where the lighter stau is the LSP, 
 while the region $m_{3/2}$ $\lesssim$ 20 TeV is excluded 
 by the current lower bounds on chargino and Higgs masses. 

The corresponding result for $M_5$ $=$ 500 TeV is presented in Fig. 2.  
The dotted, solid, dot--long-dashed short-dash--long-dashed lines 
 correspond to $\Omega_{\chi} h^2$ $=$ 0.5, 0.1, 0.01 and 0.005, respectively.
It is seen that the relic density is greatly enhanced due to 
the increase of the Hubble expansion rate in brane world cosmology. 
The allowed region (shaded area) appears for 
100 TeV $\lesssim$ $m_{3/2}$ $\lesssim$ 110 TeV. 
In this region, the neutralino mass is around 300 GeV. 

The result for $M_5$ $=$ 100 TeV is shown in Fig. 3. 
The relic density is further enhanced, and the allowed region is shifted
to lie around 40 TeV $\lesssim$ $m_{3/2}$ $\lesssim$ 45 TeV. 
In this allowed region, the lightest neutralino mass is around 130 GeV. 

These figures clearly indicate that, 
 as $M_5$ decreases to $\mathcal{O}(100)$ TeV, 
 the allowed region moves to the left and 
 the mass of the W-ino-like neutralino becomes small. 
As $M_5$ decreases further, 
 the allowed region eventually disappears. 

Figures 4-6 contains the similar results for $\tan\beta=40$. 
 It is seen that the contours of the relic density for $\tan\beta=40$ is 
 quite similar to that for $\tan\beta=10$. 
Distortions around $m_0 \simeq 1800$ GeV in Fig. 5 
 and $m_0 \simeq 900$ GeV in Fig. 6 
 are due to the resonance effect by 
 the CP-odd Higgs boson with mass $m_A \simeq 2 m_\chi$. 
As $M_5$ decreases, the allowed region moves to the left 
 and disappears eventually as in the case of $\tan\beta=10$. 
For $\tan\beta=40$, $b\to s\gamma$ constraint is severer than 
 $\tan\beta=10$ case, and it nearly excludes the region 
 $m_{0}$ $\lesssim$ 1 TeV and $m_{3/2}$ $\lesssim$ 100 TeV. 

For $M_5 = \mathcal{O}(100)$ TeV, 
 we can find the allowed region for W-ino dark matter in the parameter space,
while for $M_5 \lesssim 100$ TeV no allowed region appears.
This value of $M_5$ agrees with the result of rough estimation 
 in the previous section,
 and the expectation is confirmed.

\subsection{Implication to the detection}

As we have shown, in brane world cosmology 
 it is found that the W-ino-like neutralino 
 with the weak scale mass can be a candidate for cold dark matter.
In this section, toward verifying or excluding the scenario, 
 we discuss the implication to direct and indirect 
 detection experiments. 

We study the spin-independent cross section for 
 the neutralino-proton elastic scattering $\chi p \to \chi p$. 
The fundamental process for the direct detection is 
 the neutralino-quark scattering $\chi q \rightarrow \chi q$.
The relevant effective interactions in the non-relativistic limit 
are given as
\begin{equation}
\mathcal{L} =
 d(\bar{\chi}\gamma^{\mu}\gamma_5\chi)(\bar{q}\gamma_{\mu}\gamma_5 q)
 + f(\bar{\chi}\chi)(\bar{q}q).
\label{DetectionInteraction}
\end{equation}
The interaction can be divided into two part, the spin-dependent part 
 which consists 
 of the t-channel Z boson exchange and the s-channel squark exchange, and 
 the spin-independent part which consists of the t-channel neutral 
 light and heavy Higgs bosons exchange and the s-channel squark exchange.
The first term in Eq. (\ref{DetectionInteraction}) corresponds to 
 the spin-dependent interaction and 
 the second term corresponds to the spin-independent interaction.
The spin-independent cross section for 
$\chi p$ $\to$ $\chi p$ scattering, $\sigma_p^{\rm SI}$, 
can be calculated using 
the coefficient $f$ in Eq. (\ref{DetectionInteraction})
\cite{ref:direct-detection}.

Figures 7-9 contain the contour plot of the spin-independent cross section 
 in the ($m_{3/2},\, m_0$) plane for $\tan\beta =$ 10, 30, and 40. 
The five-dimensional Planck mass is fixed at $M_5 = 100$ TeV
in these figures. 
The number associated with each line represents the value of the 
cross section normalized by $10^{-9}$ pb.
In the shaded region, the WMAP constraint is satisfied. 

For $\tan\beta =10$ (Fig. 7), 
 the expected detection cross section consistent with 
 the WMAP result has an upper bound 
 as $\sigma_p^{\rm SI} \lesssim  5\times 10^{-9}$ pb, 
 and the cross section decreases as $m_{3/2}$ increases.  
For $\tan\beta =30$ (Fig. 8),  
the detection cross section is slightly enhanced compared with
$\tan\beta =10$ case, and 
$\sigma_p^{\rm SI}$ can be larger than $5\times 10^{-9}$ pb 
in the WMAP allowed region. 
For $\tan\beta =40$ (Fig. 9),  
the detection cross section is much more enhanced, and 
$\sigma_p^{\rm SI}$ can be as large as $10^{-7}-10^{-8}$ pb in the region
consistent with the WMAP result. 

Comparing with Figs 1-6, one may see 
 the corresponding allowed region of $\Omega_{\chi}h^2$ for another $M_5$.

As an indirect search of W-ino dark matter in the universe, 
 we can consider fluxes of cosmic positrons, anti-protons, 
 and gamma-rays originating from the annihilation 
 of dark matter in the galactic halo. 
Interestingly, the HEAT experiment reported a flux of cosmic positrons 
 in excess of the predicted rate, peaking around 10 GeV \cite{HEAT}. 
This would be reasonably explained by the annihilation 
 of the W-ino-like neutralino with mass around 300 GeV \cite{HS}. 
Upcoming experiments such as PAMELA \cite{PAMELA} and AMS-02 
 may be able to prove the excess of the positron flux  
 and probe the W-ino-like neutralino with the weak scale mass.

\section{Conclusions}

We have studied the W-ino-like neutralino relic density 
 with the mass spectrum of AMSB model
 in the context of brane world cosmology.
Although the W-ino-like neutralino with the weak scale mass 
 cannot be a candidate for dark matter 
 in the ``standard cosmology'' because of too efficient annihilation 
 under the assumption that the dark matter is a thermal relic,
 the modification of the expansion law in brane world cosmology, 
 namely the existence of $\rho^2$ term, 
 can enhance the relic density of dark matter, 
 so that the W-ino-like neutralino with mass of order $100$ GeV 
 can be a reasonable candidate for cold dark matter. 
We have performed numerical calculations and 
 shown that the WMAP allowed region moves to the left 
 in the ($m_{3/2},\, m_0$) plane 
 and the mass of the W-ino-like neutralino becomes small 
 to the order of 100 GeV, 
 as the five-dimensional Planck mass $M_5$ is decreasing. 
The allowed region disappears eventually, 
 when $M_5$ becomes as low as 100 TeV. 
While it has been reported that the lower bound on $M_5$ is 
 $M_5 \gtrsim 600$ TeV for neutralino dark matter 
 in the constrained MSSM \cite{NOS},
 this bound is moderated as $M_5 \gtrsim 100$ TeV 
 for the case of the W-ino-like neutralino. 
In addition, we have briefly discussed the direct and indirect 
 detections of the W-ino-like neutralino with mass of order $100$ GeV.

%
\section*{ACKNOWLEDGMENTS}
We would like to thank Shigeki Matsumoto for useful discussions. 
The works of T.N. and N.O. are supported in part 
 by the Grant-in-Aid for Scientific Research (\#16740150 and  \#15740164) 
 from the Ministry of Education, Culture, Sports, 
 Science and Technology of Japan.  
The work of O.S. is supported by PPARC.




\begin{figure}[p]
\begin{center}
\epsfig{file=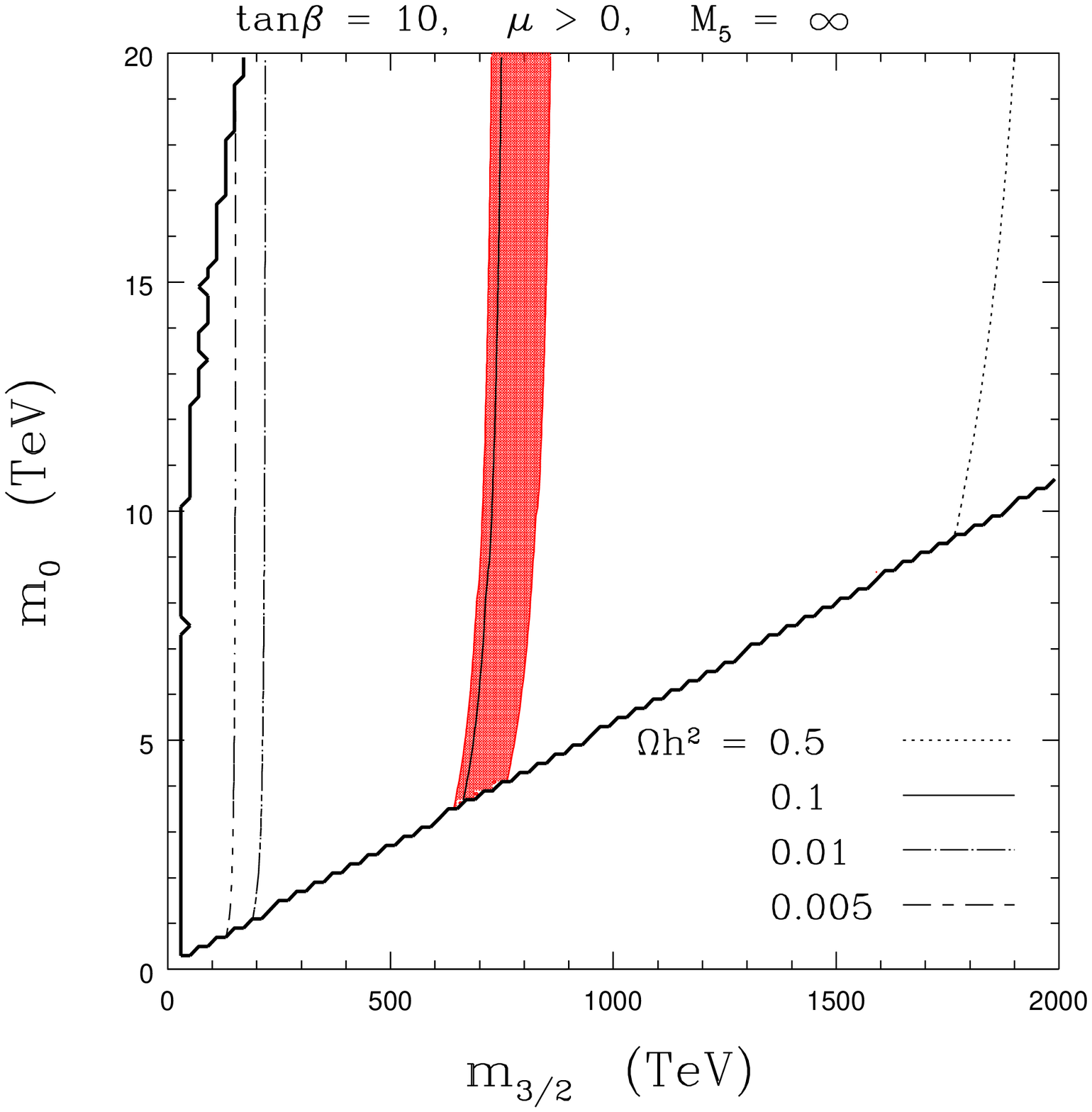,width=8cm}
\end{center}
\caption{\label{Fig1a}
Contours of the relic density $\Omega_{\chi}h^2$ in the 
($m_{3/2},\, m_0$) plane for $\tan\beta = 10$, $\mu>0$ and $M_5$ $=$ $\infty$. 
The region along the bold line and the two coordinate axes 
 is excluded by various experimental constraints 
 (the lightest Higgs mass bound, $b\to s\gamma$ constraint, 
 the lighter chargino mass bound etc.)
 or the condition for the successful electroweak symmetry breaking. 
}
\end{figure}
\begin{figure}[p]
\begin{center}
\epsfig{file=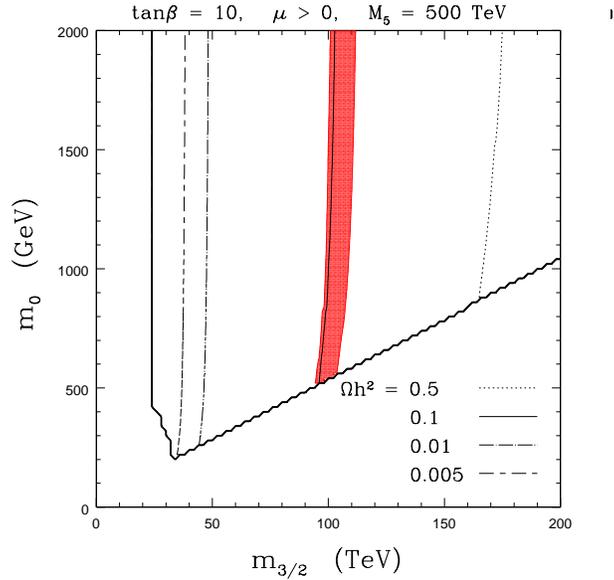,width=8cm}
\end{center}
\caption{\label{Fig1b}
Contours of the relic density $\Omega_{\chi}h^2$ in the 
($m_{3/2},\, m_0$) plane for $\tan\beta = 10$, $\mu>0$ and 
$M_5$ $=$ 500 TeV. 
The shaded region is allowed by 
the WMAP constraint $0.094 < \Omega_{\chi} h^2 < 0.129$. 
}
\end{figure}
\begin{figure}[p]
\begin{center}
\epsfig{file=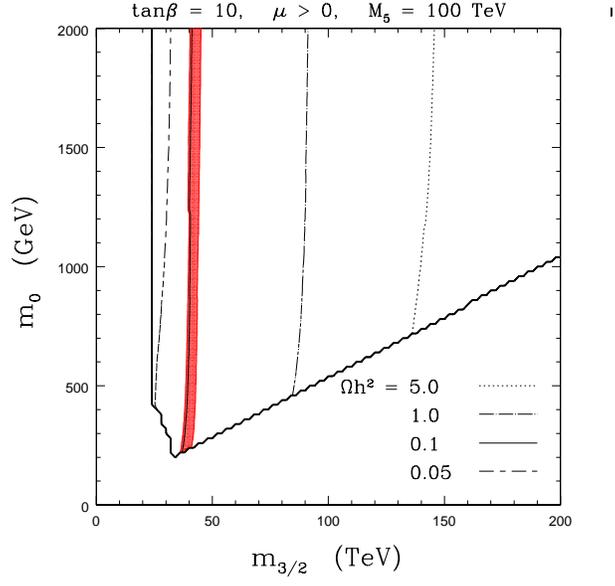,width=8cm}
\end{center}
\caption{\label{Fig1c} 
The same as Fig. 2 but for $M_5=$ 100 TeV. 
}
\end{figure}

\begin{figure}[p]
\begin{center}
\epsfig{file=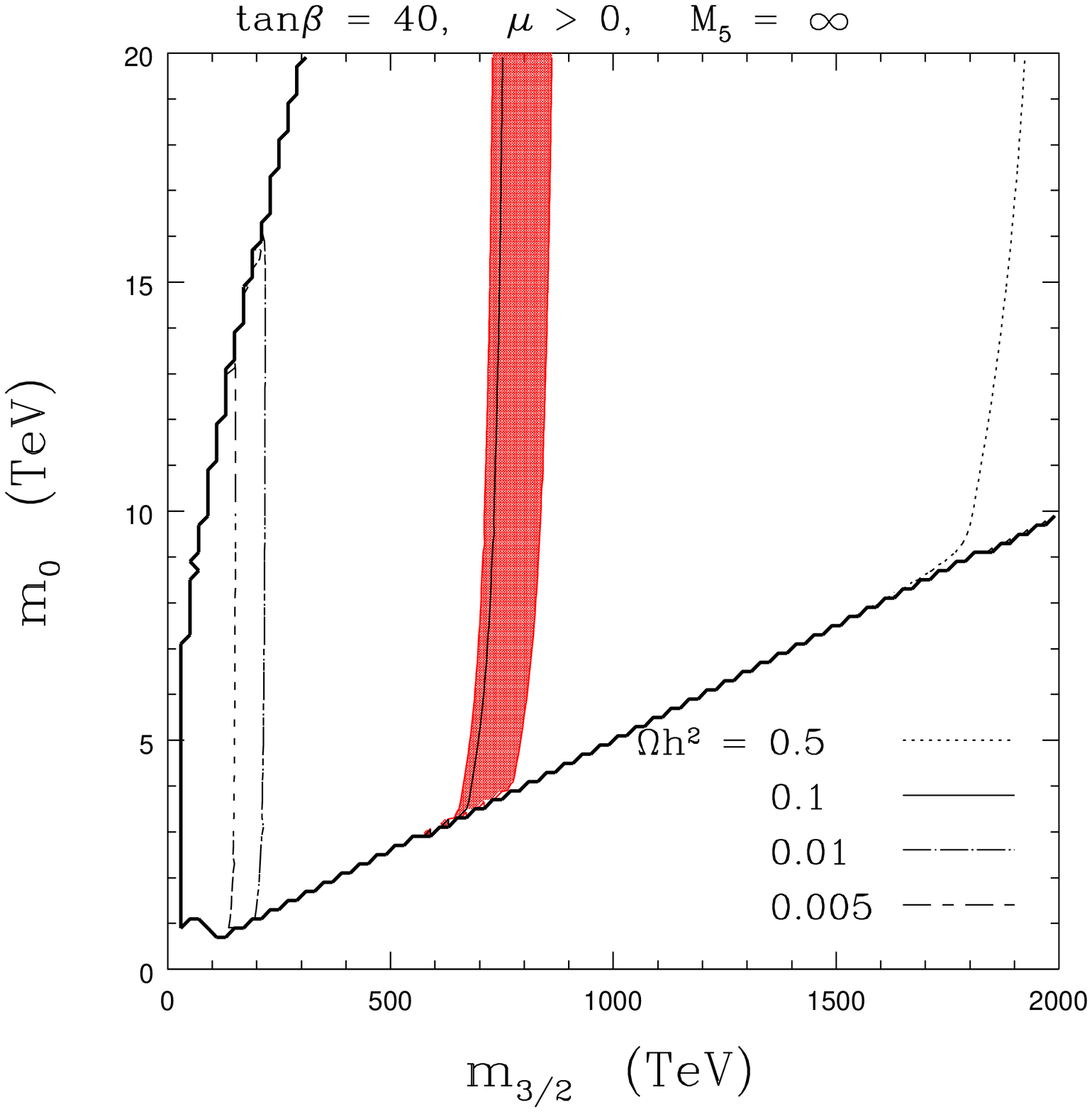,width=8cm}
\end{center}
\caption{\label{Fig2a} 
The same as Fig. 1 but for $\tan \beta = 40$. 
}
\end{figure}
\begin{figure}[p]
\begin{center}
\epsfig{file=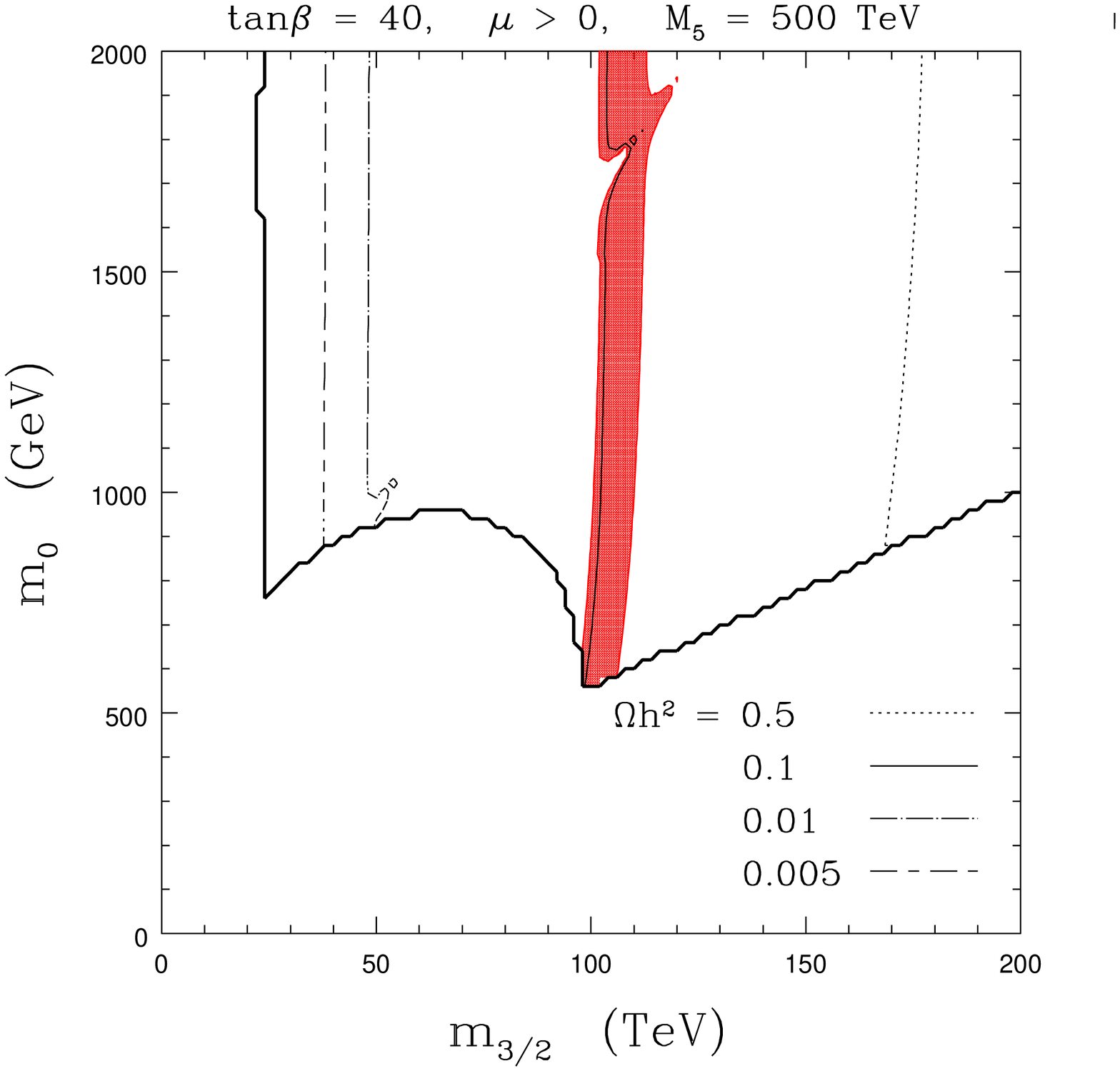,width=8cm}
\end{center}
\caption{\label{Fig2b}
The same as Fig. 2 but for $\tan \beta = 40$. 
}
\end{figure}
\begin{figure}[p]
\begin{center}
\epsfig{file=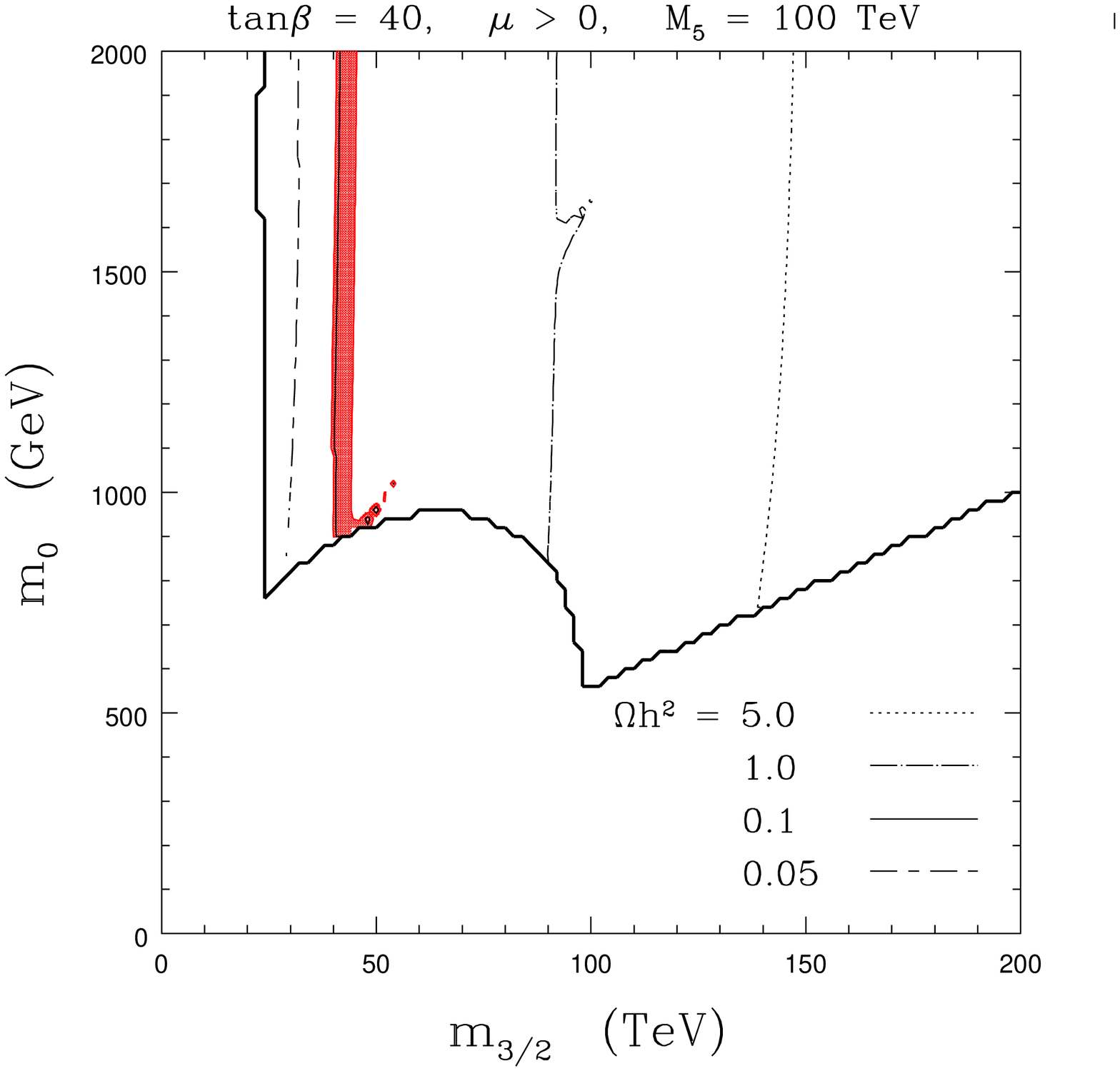,width=8cm}
\end{center}
\caption{\label{Fig2c}
The same as Fig. 3 but for $\tan \beta = 40$.  
}
\end{figure}

\newpage
\begin{figure}[p]
\begin{center}
\epsfig{file=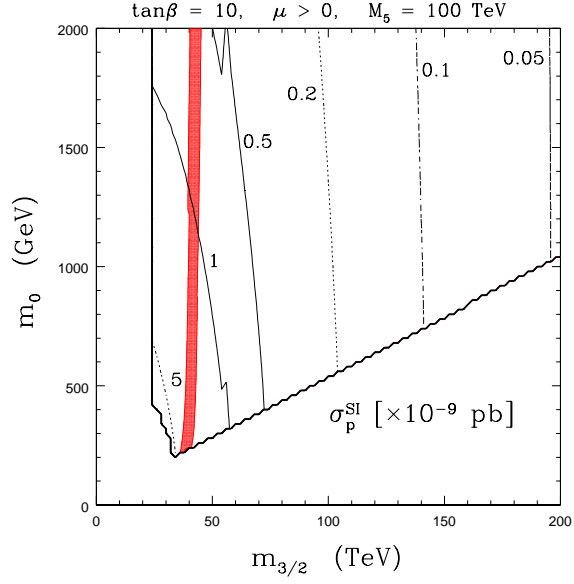,width=8cm} 
\end{center}
\caption{
\label{Fig3a}
Contours of the spin-independent cross section of $\chi p$ $\to$ $\chi p$,
$\sigma_p^{\rm SI}$, in the ($m_{3/2},\, m_0$) plane 
for $\tan\beta = 10$, $\mu>0$ and $M_5$ $=$ 100 TeV. 
The region consistent with 
the WMAP 2$\sigma$ allowed range is shaded.
}
\end{figure}
\begin{figure}[p]
\begin{center}
\epsfig{file=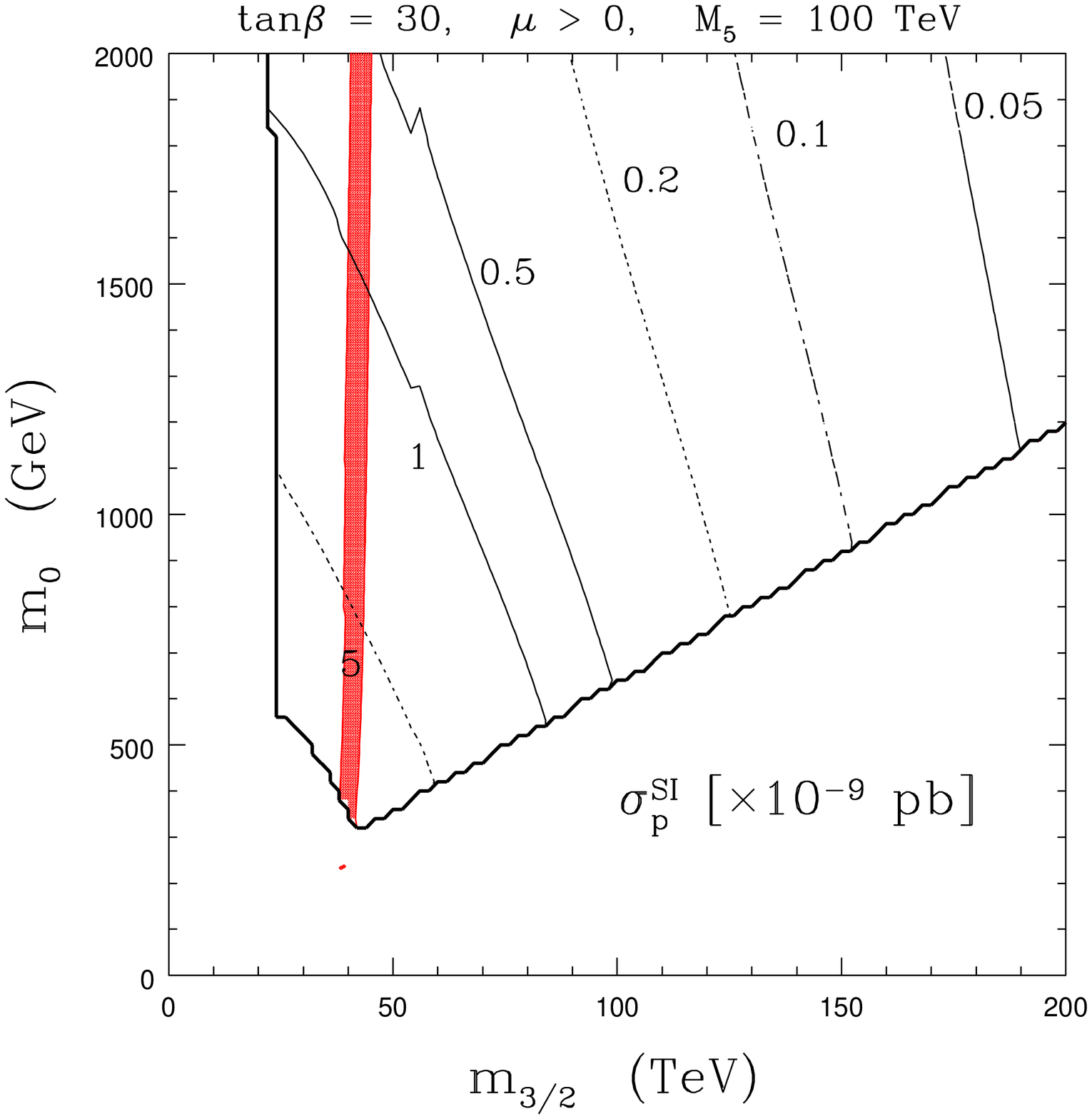,width=8cm}
\end{center}
\caption{
\label{Fig3b}
The same as Fig. 7 but for $\tan \beta = 30$.  
}
\end{figure}

\begin{figure}[p]
\begin{center}
\epsfig{file=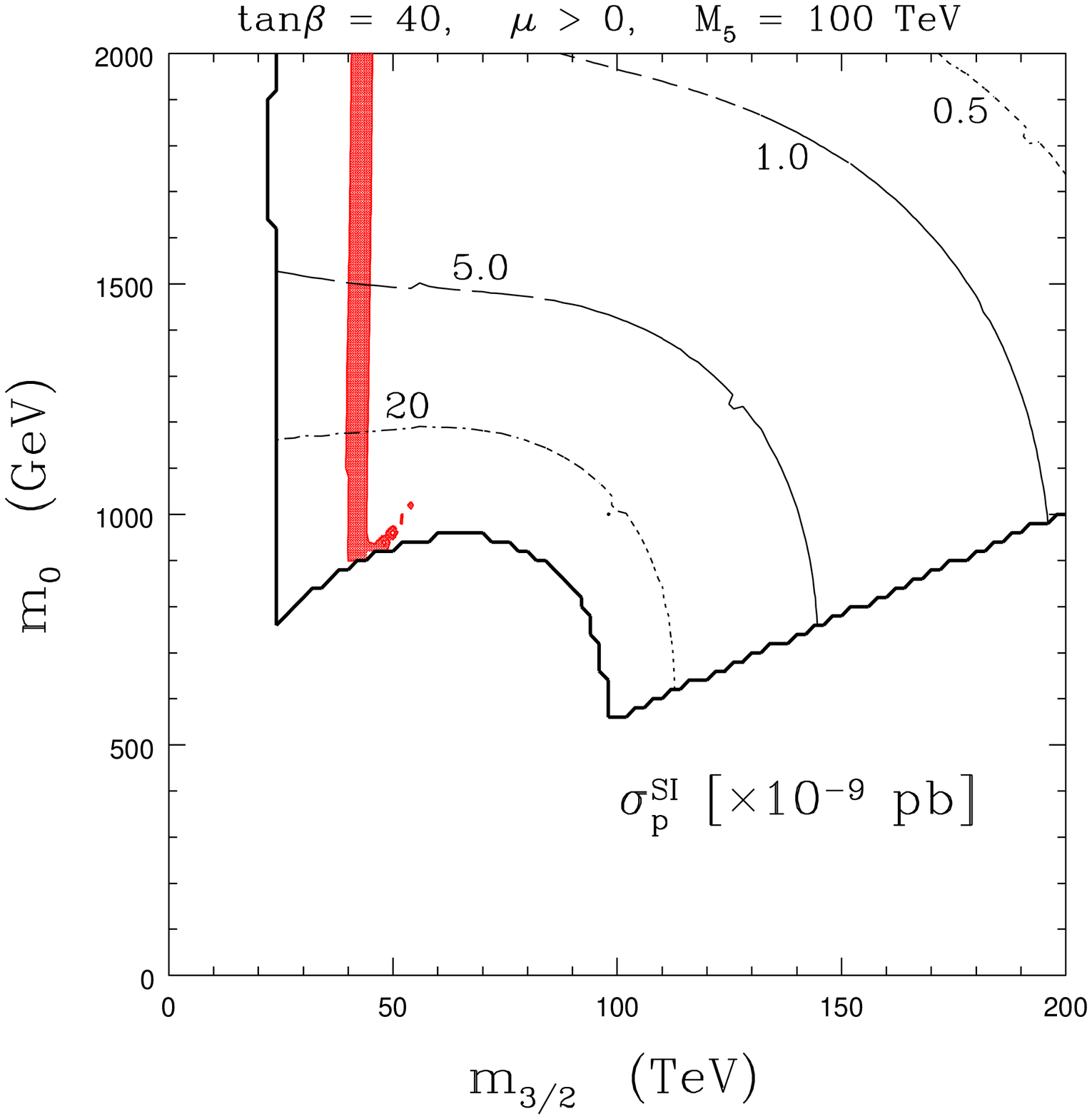,width=8cm}
\end{center}
\caption{
\label{Fig3c}
The same as Fig. 7 but for $\tan \beta = 40$.  
}
\end{figure}



\begin{thebibliography}{99}

\bibitem{WMAP}
D.~N.~Spergel {\it et al.}, 
 Astrophys.\ J.\ Suppl.\ {\bf 148}, 175 (2003).

\bibitem{CDM} 
See, for example, 
J.~R.~Ellis, K.~A.~Olive, Y.~Santoso and V.~C.~Spanos,
 Phys.\ Lett.\ B {\bf 565}, 176 (2003); 
A.~B.~Lahanas and D.~V.~Nanopoulos,
 Phys.\ Lett.\ B {\bf 568}, 55 (2003).

\bibitem{AnomalyMediation1}
L.~Randall and R.~Sundrum,
 Nucl.\ Phys.\ B {\bf 557}, 79 (1999).

\bibitem{AnomalyMediation2}
G.~F.~Giudice, M.~A.~Luty, H.~Murayama and R.~Rattazzi,
 JHEP {\bf 9812}, 027 (1998).

\bibitem{Mizuta}
S.~Mizuta, D.~Ng and M.~Yamaguchi, Phys.\ Lett.\ B {\bf 300}, 96 (1993).

\bibitem{MoroiRandall}
T.~Moroi and L.~Randall, Nucl.\ Phys.\ B {\bf 570}, 455 (2000).

\bibitem{FujiiHamaguchi}
M.~Fujii and K.~Hamaguchi, Phys.\ Rev.\ D {\bf 66}, 083501 (2002).

\bibitem{Pallis}
C.~Pallis, Astropart.\ Phys.\  {\bf 21}, 689 (2004).

\bibitem{AdditionalScalar}
S.~Profumo and P.~Ullio, JCAP {\bf 0311}, 006 (2003); 
C.~Pallis, JCAP {\bf 0510}, 015 (2005).

\bibitem{braneworld}
For a review, see, e.g.,
D.~Langlois, Prog.\ Theor.\ Phys.\ Suppl.\ {\bf 148}, 181 (2003).

\bibitem{RS}
L.~Randall and R.~Sundrum,
 Phys.\ Rev.\ Lett.\ {\bf 83}, 4690 (1999).

\bibitem{maedawands}
K.~i.~Maeda and D.~Wands,
 Phys.\ Rev.\ D {\bf 62}, 124009 (2000).

\bibitem{OS} 
N.~Okada and O.~Seto,
 Phys.\ Rev.\ D {\bf 70}, 083531 (2004).

\bibitem{NOS} 
T.~Nihei, N.~Okada and O.~Seto, 
 Phys.\ Rev.\ D {\bf 71}, 063535 (2005).

\bibitem{OSGravitino} 
N.~Okada and O.~Seto, 
 Phys.\ Rev.\ D {\bf 71}, 023517 (2005).

\bibitem{Axino}
G.~Panotopoulos, JCAP {\bf 0508}, 005 (2005).

\bibitem{OSleptogenesis}
N.~Okada and O.~Seto, Phys.\ Rev.\ D {\bf 73}, 063505 (2006) ;
M.~C.~Bento, R.~Gonzalez Felipe and N.~M.~C.~Santos, 
 Phys.\ Rev.\ D {\bf 73}, 023506 (2006).

\bibitem{Kolb}
See, e.g.,
E.~W.~Kolb and M.~S.~Turner, 
 \textit{The Early Universe}, Addison-Wesley (1990).

\bibitem{ref:darksusy} 
P.~Gondolo, J.~Edsjo, L.~Bergstrom, P.~Ullio and E.A.~Baltz,
 JCAP {\bf 0407}, 008 (2004). 

\bibitem{ref:minimal-AMSB} 
T.~Gherghetta, G.F.~Giudice and J.D.~Wells, 
Nucl.\ Phys.\ B {\bf 559}, 27 (1999).

\bibitem{ref:suspect} 
A.~Djouadi, J.~L.~Kneur and G.~Moultaka, hep-ph/0211331.

\bibitem{ref:direct-detection} 
G.~Jungman, M.~Kamionkowski and K.~Griest, 
 Phys.\ Rep.\ {\bf 267}, 195 (1996).

\bibitem{HEAT}
See, for example,  
S.~Coutu {\it et al.}, 
Astropart.\ Phys.\  {\bf 11}, 429 (1999), 
  references therein. 

\bibitem{HS}
D.~Hooper and J.~Silk, 
 Phys.\ Rev.\ D {\bf 71}, 083503 (2005). 

\bibitem{PAMELA}
M.~Circella  [PAMELA Collaboration],
 Nucl.\ Instrum.\ Meth.\ A {\bf 518}, 153 (2004).


\end{thebibliography}
\end{document}